\documentclass[useAMS,usenatbib]{mn2e}

\newcommand{\xmmn}{\mbox{\em XMM-Newton}}

\newcommand{\etal}{\mbox{et\ al.\ }}

\usepackage{rotating}
\usepackage{url}
\usepackage{amsmath}
\usepackage{amssymb}

\title[The story of RE\,J2248-511]
{The story of Seyfert galaxy RE\,J2248-511: from intriguingly ultrasoft to unremarkably average}

\author[Starling, Done \& Jin \etal]
{R. L. C. Starling$^{1}$\thanks{mailto:rlcs1@le.ac.uk}, C. Done$^{2}$, C. Jin$^{2}$, E. Romero-Colmenero$^{3}$, S. B. Potter$^{3}$, 
\and K. Wiersema$^{1}$, K. L. Page$^{1}$, M. J. Page$^{4}$, A. A. Breeveld$^{4}$ and A. P. Lobban$^{1}$\\   
$^{1}$ Department of Physics and Astronomy, University of Leicester, University Road, Leicester LE1 7RH\\
$^{2}$ Department of Physics, University of Durham, South Road, Durham DH1 3LE\\
$^{3}$ South African Astronomical Observatory, P.O. Box 9, Observatory 7935, Cape Town, South Africa\\
$^{4}$ Mullard Space Science Laboratory, University College London, Holmbury St. Mary, Dorking, Surrey RH5 6NT\\   
}

\begin{document}
\date{Accepted . Received ; in original form }

\pagerange{\pageref{firstpage}--\pageref{lastpage}} \pubyear{}

\maketitle

\label{firstpage}


\begin{abstract}
RE\,J2248-511 is one of only 14 non-blazar AGN detected in the far ultraviolet by the {\it ROSAT} Wide Field Camera implying a large ultrasoft X-ray flux. This soft X-ray excess is strongly variable on year timescales, a common property of Narrow Line Seyfert 1s, yet its optical linewidths classify this source as a broad-lined Seyfert 1. We use four nearly simultaneous optical--X-ray SEDs spanning 7 years to study the spectral shape and long term variability of RE\,J2248-511.

Here we show that the continuum SED for the brightest epoch dataset is
consistent with the mean SED of a standard quasar, and matches well to
that from an XMM-SDSS sample of AGN with $\langle M/M_{\odot}
\rangle~\sim 10^8$ and $\langle L/L_{\rm Edd} \rangle~\sim 0.2$. All
the correlated optical and soft X-ray variability can be due entirely
to a major absorption event. The only remarkable aspect of this AGN is
that there is no measurable intrinsic X-ray absorption column in the
brightest epoch dataset. The observed FUV
flux is determined by the combination of this and the fact that the
source lies within a local absorption `hole'. RE\,J2248-511, whose variable, 
ultrasoft X-ray flux once challenged its BLS1 classification, demonstrates that characterisation of such
objects requires multi-epoch, multi-wavelength campaigns.
\end{abstract}


\section{Introduction}

The ultraviolet (UV) to X-ray spectral shapes seen in active galactic
nuclei (AGN) clearly comprise multiple components and have been the
subject of decades of study. There is an accretion disc, which peaks
in the UV (forming the Big Blue Bump) in the standard optically thick,
geometrically thin disc solutions \citep{SS73}. This emission forms
the seed photons for Compton upscattering in a hot corona, producing a
power law above 2\,keV. However, an excess of soft ($< 2$\,keV) X-ray
emission above this power law emission is seen ubiquitously in all
high mass accretion rate AGN. The origin of this soft X-ray excess
component is currently unknown but if this represents a true
additional continuum \citep[e.g.][]{Laor,Magdziarz,GierDone04}, then it has very similar temperature in all
sources \citep{Czerny,GierDone04,Middleton}. This apparent fine tuning led to alternative
models where the soft excess instead arises via reflection/absorption
from partially ionised material, so that atomic features set the fixed
energy \citep{Crummy,GierDone04,Middleton,Walton}. However, both these scenarios predict strong
atomic features, at odds with the observed smooth soft X-ray excess
continuum. These features can be smoothed into a pseudo-continuum by
strong velocity shear, but the velocities required appear too extreme to
be plausible for an absorbing wind \citep{Schurch}, and even in
reflection require the very innermost edge of the disc around a high
spin black hole \citep{Crummy,Walton}. Reflection
also has the problem that the ionisation state must be fine-tuned
\citep{DoneNayakshin}, resurrecting the very issue that the model
was designed to avoid.

The narrow-line Seyfert 1 (NLS1) subclass, objects where the FWHM of the broad component of the H$\beta$ line is $\le 2000$ km s$^{-1}$ \citep{OP}, show nearly ubiquitous strong, steep soft X-ray emission \citep{BBF,Gru}. These are most probably low mass black holes accreting at high mass accretion rates with respect to Eddington \citep[$L/L_{\rm Edd} \sim 1$,][]{Boroson}, so much of the soft excess may well be due to intrinsic emission from the accretion disc (Done et al. 2012, hereafter D12; Jin et al. 2012a, hereafter J12a).
The combination of low mass and high $L/L_{\rm Edd}$ predict that
the NLS1 should have the highest intrinsic accretion disc temperature
of all AGN, and that this can peak close to (or even in) the soft
X-ray bandpass. These objects do still need an additional soft X-ray
component `filling in' smoothly between the disc and high energy power
law, but this `true' soft X-ray excess is a much smaller fraction of
the inferred bolometric luminosity than in standard Broad Line Seyfert
1's (D12; J12a). 

A subset of NLS1 also show dramatic variability, with deep dips in
X-ray flux. These dip spectra are extremely complex \citep{Gallo}, but
can be fit with extremely smeared relativistic reflection \citep[e.g.][]{Fabian09,Fabian13}. However, the NLS1 which do not show such dips tend
to have rather simple spectra (see Middleton et al. 2009, Jin et al
2009, J12a, and Jin, Ward \& Done 2012c). In these `simple'
sources, the fast variability strongly favours the original model of a
separate soft X-ray Comptonisation component \citep{Jin2013}. Standard broad
line Seyferts also show growing evidence from variability studies for a
separate soft X-ray component \citep[e.g.][]{Meh,Noda11,Noda,Lohfink}. While the fine-tuning of the
temperature is still an issue, such a component could arise from shock
heating of the disc surface from the impact of a failed wind \citep{Done13}.

Many AGN samples have been studied through their SEDs, facilitated by
large extragalactic surveys undertaken with e.g. {\it XMM-Newton},
{\it Chandra} and the Sloan Digital Sky Survey (SDSS) and by the
availability of simultaneous otical to X-ray datasets from {\it
XMM-Newton} and {\it Swift} which avoid any confusion introduced by
variability \citep[e.g.][]{Brocksopp,VandF09,Gru10}. While this has highlighted the diversity in
broadband spectral shapes and their variability, it has also been
possible to create mean spectra for particular classes (e.g. Elvis et
al. 1994; J12a). A number of SED studies have concluded
that Eddington ratio is a main driver for the SED shape
(e.g. Vasudevan \& Fabian 2007,2009; J12a). If this is true then the physics responsible for the soft X-ray excesses in NLS1 is likely to be closely tied to the Eddington ratio. Hence
defining this class of objects on a single line width measure would then be inappropriate, missing higher mass black holes at similarly high
$L/L_{\rm Edd}$ \citep{Collin,Sulentic,Dultzin,Peterson}. However, high mass,
high $L/L_{\rm Edd}$ AGN are rare in the local Universe, predominantly due
to downsizing of activity in the Universe from $z \sim 2$, so the single
line width criterion may not miss many local objects. 

RE\,J2248-511 is a local ($z\sim0.1$), EUV-selected Seyfert galaxy. It is classified as a broad-line Seyfert 1 (BLS1) due to its Balmer broad-line widths of FWHM$\sim$3600 km s$^{-1}$.
For an estimated black hole mass of $\sim$10$^{8}$\,M$_{\odot}$, this source is highly variable in both optical and soft X-rays \citep{emp95,aad}. {\it ROSAT} measured a soft X-ray spectral slope of $\Gamma \sim 3$ and it was among the tiny fraction of {\it ROSAT} All Sky Survey (RASS) AGN also detected, and in this case discovered, by the {\it ROSAT} Wide Field Camera in the far ultraviolet \citep{Po}. Despite its soft X-ray slope, there is no evidence for strong {\small Fe\,II} emission \citep[e.g.][]{Wilkes,BBF}. This curious mixture of properties often places RE\,J2248-511 in the crossover of parameter space between BLS1 and NLS1.
 
RE\,J2248-511 is then an object where we can examine the nature of the
soft X-ray excess and the relationship between the X-ray and
optical/UV continua. To probe the underlying physics of this AGN we
have gathered four epochs of quasi-simultaneous optical and X-ray
observations using the SAAO 1.9m Radcliffe telescope, the Danish 1.54m
telescope at La Silla, {\it XMM-Newton} and {\it Swift}, spanning 7
years from 2000 October to 2007 September. We also make a comparison
to archival data. Section \ref{sec:obs} describes the observations and
in Section \ref{sec:bhmass} we confirm the broad-line cloud velocities
and estimate the black hole mass from our optical spectroscopy. We
then determine whether the SEDs can be well represented by the
broadband spectral models of D12 in Section \ref{sec:modelling}, and
evaluate spectral variability between epochs. We discuss our findings
and conclude in Section \ref{sec:discussion}.

\begin{table*}
\begin{center}
\caption{UV, Optical and X-ray observations utilised in this work.}
\begin{tabular}{lllll} 
Waveband & Obs. date& Telescope and Instrument & T$_{\rm exp}$ (s) & Reference\\ \hline \hline
X-ray & 2007 September 26& {\it Swift} XRT &1919 & Grupe et al. (2010)\\
& 2007 September 15& {\it Swift} XRT & 22569&Grupe et al. (2010)\\
 & 2007 May 15-16 & {\it XMM-Newton} EPIC pn & 45058 & Dunn et al. (2010)\\
 & 2007 May 15 & {\it XMM-Newton} EPIC MOS 1,2 &  57579,58848 & Dunn et al. (2010)\\
& 2006 November 01 & {\it Swift} XRT & 5769 & Grupe et al. (2010)\\
& 2006 September 21-22 & {\it Swift} XRT & 9759&  Grupe et al. (2010)\\
 & 2001 October 31 & {\it XMM-Newton} EPIC pn & 9767 & \\
 & 2001 October 31 & {\it XMM-Newton} EPIC MOS 1,2 & 14301,14304 &\\
 & 2000 October 26 & {\it XMM-Newton} EPIC pn & 10089 &\\
 & 2000 October 26 & {\it XMM-Newton} EPIC MOS 1,2 & 13789,13791 &\\
 & 1997 May 17 & {\it ASCA} GIS2,GIS3 & 19960,19956 & Breeveld et al. (2001)\\
 & 1993 April 21&{\it ROSAT} PSPC & 4520 & Pounds et al. (1993); Puchnarewicz et al. (1995)\\ \hline
UV & 2004 July 07 & {\it FUSE} & 3299  &Dunn et al. (2010)\\
   & 2002 September 24 & {\it FUSE} & 5534 &Dunn et al. (2010)\\
  & 2002 September 24 & {\it FUSE} & 31301 &Dunn et al. (2010)\\
   & 1992 November 23 & {\it IUE} SWP & 10799.8 & Dunn et al. (2006)\\ \hline
optical  & 2007 May 16 & SAAO 1.9m (low-res blue,red) & 900&\\
 & 2006 August 29 & Danish 1.54m DFOSC B,V,R&180,2x120,60 &\\
 & 2001 October 15 & SAAO 1.9m (blue grating)& 900&\\
 & 2000 October 19-20 & SAAO 1.9m (low-res blue,red) & 500,500&\\
 & 2000 October 19-20 & SAAO 1.9m (high-res blue,red) & 500,500&\\
 & 2000 October 21-22 &SAAO 1.9m (low-res blue,red) & 600,900&\\
 & 2000 October 21-22 &SAAO 1.9m (high-res blue,red) & 600,600&\\
 & 2000 October 23-24 & SAAO 1.9m (low-res blue,red) & 600,900&\\
 & 2000 October 23-24 & SAAO 1.9m (high-res blue,red) & 600,600&\\
\label{tab:obs}
\end{tabular}
\end{center}
\end{table*}

\section{Observations} \label{sec:obs}

\subsection{X-ray}
RE\,J2248-511 was observed with \emph{XMM-Newton} \citep{Jan} on 2000 October 26 (observations 0109070401 and 0109070501), 2000 October 31 (0109070601) and 2007 May 15 (0510380101) as listed in Table \ref{tab:obs}. We do not use observation 0109070501 due to sustained background flaring. All EPIC \citep{Stru} instruments were in small window mode with the medium filter
applied, excepting the MOS cameras during the 2000 observation which were then in small window free running mode.
The raw data were processed with the \emph{XMM} SAS version 20110223\_1801-11.0.0.

The source was observed with the X-Ray Telescope \citep[XRT,][]{XRT} on board
\emph{Swift} \citep{Swift} on 2006 September 21--22, 2006 November 1, and on 2007 September 15 and 26 (Table \ref{tab:obs}). We obtained the extracted spectra from the UK Swift Science Data Centre\footnote{\url{www.swift.ac.uk/user_objects}}, combining the two 2007 observations into a single spectrum following the procedures of \cite{Evans}.
Data were processed with the \emph{Swift} software version 3.9 using CALDB 3.9.

All X-ray spectra were grouped such that a minimum of 20 counts fell in each bin, and we used the energy range 0.3--10\,keV for analysis of {\it XMM-Newton} and {\it Swift} X-ray spectra.

\subsection{Optical}
A bright star ($\epsilon$ Gru with $V \sim 3.5$) is
located close ($\sim$9.1 arcmin) to the AGN, which made observations with the
\xmmn\ Optical Monitor and the \emph{Swift} UV Optical Telescope impossible. Observations were, however, possible with ground-based telescopes, listed in Table \ref{tab:obs}.

RE\,J2248-511 was observed on
2000 October 19--24, 2001 October 15 and 2007 May 16 using the South African
Astronomical Observatory's 1.9m Radcliffe telescope with the grating-spectrograph and SITe CCD. These observations coincide with the \xmmn\ observations to within 2--6 days in 2000, 2.5 weeks in 2001 and 2 days in 2007. Spectra were taken using both a narrow slit ($\sim$1.8$''$) and a much wider slit,
with overlapping blue (\#7) and red (\#8) gratings ($\lambda_{\rm central}$ = 4600, 7800\AA) in 2000 and 2007. Only the blue grating was used in 2001. Exposure times ranged from 500\,s to 900\,s per spectrum. Arc spectra were taken before and
after every target and every standard spectrum using a CuAr lamp. The standard
star LTT\,9239 was observed for flux calibration of RE\,J2248-511. 
Spectra were reduced using standard tools within {\small IRAF}. In 2000 spectra were obtained on 3 separate nights and under variable weather conditions, during which no appreciable variability is seen. For continuum measurement and SEDs in 2000 October we use the spectra on 21st--22nd which were taken under the best seeing conditions. We use the wide-slit data for SED modelling and the narrow-slit data for spectral line measurements.

We observed RE\,J2248-511 on 2006 August 29 with the Danish 1.5\,m telescope at La Silla,
using the DFOSC instrument for optical photometry. These observations lie within 3 weeks of the first {\it Swift} pointing.
The following filters and exposure times were used: 180 seconds in $B$, $2 \times 120$ seconds in $V$ and
60 seconds in $R$. 
We chose the position of the AGN on the chip of
DFOSC and the timing of the observations such that the bright star $\epsilon$ Gru
is off the chip and diffraction spikes are not affecting the AGN. The data
were reduced using standard procedures in {\small IRAF}, using
sky flats and bias frames taken at the end of the night. Photometric
calibrations of the $B$ and $V$ data were performed using magnitudes of field stars from the
AAVSO Photometric All-Sky Survey (APASS) survey\footnote{http://www.aavso.org/apass} data release 6.
To calibrate the $R$ band we used a transformation of APASS Sloan $r'$ and $i'$ magnitudes to $R$ following \cite{Jordi}.
We find the following magnitudes: $B = 15.55 \pm 0.07$,
$V = 15.45 \pm 0.05$ and $R = 15.29 \pm 0.06$, approximately equivalent to (3.9$\pm$0.3,2.3$\pm$0.1,1.46$\pm$0.08)$\times$10$^{-15}$ erg cm$^{-2}$ s$^{-1}$ \AA$^{-1}$ respectively. 

\section{The black hole mass} \label{sec:bhmass}
\begin{figure}
\begin{center}
\includegraphics[width=6cm, angle=0]{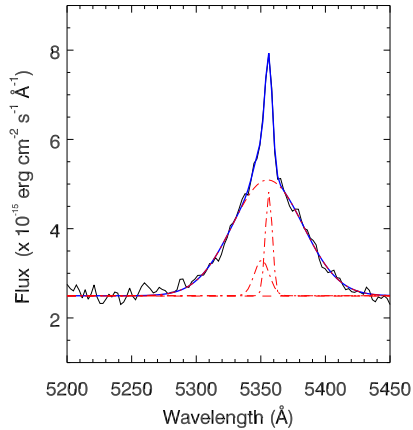}
\caption{Fit to the observed H$\beta$ profile during the 2000 October observations at SAAO.}
\label{fig:hbeta}
\end{center}
\end{figure}

The optical spectra show strong emission lines from H$\alpha$,$\beta$,$\gamma$,$\delta$, $[$O\,II$]$,$[$O\,III$]$, $[$N\,II$]$ and $[$Ne\,III$]$. For this work we concentrate on the continuum and hydrogen line fits, in order to estimate the black hole mass.
Optical spectral fitting was done with both the Starlink {\small DIPSO} v3.6 spectral fitting package and {\sc IRAF}. The continuum under each line was approximated by a first order polynomial in the immediate vicinity of the line blend, Gaussian profiles were assumed for the
line profiles and a least squares procedure was used to minimise the residuals
of the fit. 
Both H$\alpha$ and H$\beta$ were best fitted with three Gaussians
representing narrow, intermediate and broad components (e.g. Figure \ref{fig:hbeta}). Towards the blue end of the spectrum the noise increased and the weaker broad lines of H$\gamma$ and H$\delta$ required only double and single Gaussian components respectively. The positions of the line centres indicate a redshift of $z = 0.1015\pm0.0010$, consistent with the best published measurement of $z = 0.1016 \pm 0.0001$ \citep{DunnFUSE08}.

\begin{table} 
\begin{center}
\caption{H$\beta$ line fit in 2000 October, requiring 3 Gaussian components. All quantities are given in the observed frame, and errors are statistical only.}
\begin{tabular}{cccc} 
$\lambda_{\rm central}$ & Amplitude & FWHM & FWHM \\     
\AA & erg cm$^{-2}$ s$^{-1}$ \AA$^{-1}$ &   \AA       & km s$^{-1}$ \\ \hline \hline
5356.2$\pm$0.2 & (2.32$\pm$0.23)$\times$10$^{-15}$  &  6.4$\pm$0.7 &  358$\pm$40 \\
5350.8$\pm$1.7 & (8.03$\pm$1.8)$\times$10$^{-16}$    & 13.7$\pm$3.1 &  768$\pm$174 \\
5355.3$\pm$0.5 & (2.59$\pm$0.06)$\times$10$^{-15}$ &  64.6$\pm$0.9 &  3619$\pm$51 \\
\label{tab:hbeta}
\end{tabular}
\end{center}
\end{table}

We use the H$\beta$ broad-line width (Table \ref{tab:hbeta} and Figure \ref{fig:hbeta}) together with either the luminosity from narrow-slit spectroscopy in 2000 October, or the continuum flux at 5100\AA\ (restframe) of 1.36$\times$10$^{-15}$ erg cm$^{-2}$ s$^{-1}$ \AA$^{-1}$ measured in the 2000 October 21/22 wide-slit data, to estimate the black hole mass. To convert flux to luminosity we used $\Omega_{\rm M}=0.27$ and H$_{0}=71$ km s$^{-1}$ Mpc$^{-1}$. We follow the methods of \cite[][their Section 3.4]{VandP} for black hole mass estimation, with both equations resulting in estimates of $\log M_{\rm BH} = 8.1$.

\section{Spectral modelling} \label{sec:modelling}
We combined the optical information with the X-ray data for our four quasi-simultaneous epochs. We removed the emission lines from the optical spectra and fitted the continua with a double power law which we use in the broadband fits \citep[e.g.][]{VandenBerk}. All modelling is performed in the X-ray spectral fitting package Xspec \citep{Xspec}. We use abundances from \cite{Wilms} and the cross-sections of \cite{Verner}. We fixed Galactic ($z=0$) X-ray absorption to 9.43$\times$10$^{19}$ cm$^{-2}$ (LAB H\,I Survey, \citealt{Kalberla}), and set reddening to $1.7\times N_{\rm H,Gal}$/10$^{22}$ \citep{Bessell} as employed in J12a (and references therein). 
We note that $N_{\rm H,Gal}$ does not change significantly with respect to the LAB value when using the new method of \cite{Willingale}, but is lower than the value of 1.4$\times$10$^{20}$ cm$^{-2}$ used in the earlier X-ray studies \cite{emp95} and \cite{aad}.

For the epochs 2000 and 2001, we fitted the X-ray data with a simple absorbed power law plus blackbody model with all parameters excepting redshift and Galactic absorption free. This provided a good fit to the overall shape of the spectra, so a constant factor was included for the MOS spectra to determine the offsets between the {\it XMM-Newton} EPIC pn and MOS. These could then be fixed in more complex physical modelelling, to 1.09 (MOS1), 1.11 (MOS2) in 2000 and 1.06 (MOS1), 1.05 (MOS2) in 2001. The same procedure was carried out for the 2007 X-ray data, but a further two blackbodies were introduced in order to adequately reproduce the spectral shape: the {\it Swift} XRT spectrum did not require an offset, MOS1 was set to 1.02 and MOS2 to 1.01 with respect to pn.
We did not allow offsets of the optical data with respect to the X-ray data.
Clearly, a combination of multiple power laws and blackbodies is sufficient to describe the spectral shape at all epochs, but this is not a physically motivated model. We therefore go on to model each epoch in turn with more viable models for the emission mechanisms in AGN, which are necessarily more complex. All plots are shown in the observer frame.

\begin{figure*}
\begin{center}
\includegraphics[width=6cm, angle=-90]{figure2a.ps}
\includegraphics[width=6cm, angle=-90]{figure2b.ps}
\caption{This figure illustrates the colour temperature corrected accretion disc plus Comptonisation models ({\it optxagnf}, left) and including reflection ({\it optxagnf+(rdblur*pexmon)}, right), which we applied to the SEDs. The black curves show the total unabsorbed model from the best fit to the epoch 2001 SED: red curves show the disc component; green curves show the soft X-ray Compton component; blue curves show the hard X-ray Compton component; the orange curve shows the {\it rdblur*pexmon} reflection component (which includes an Fe line, visible here only in the total spectrum). The y-axis is in units keV$^2$ photons cm$^{-2}$ s$^{-1}$ keV$^{-1}$.}
\label{Modelcomponents}
\end{center} 
\end{figure*}

\begin{figure}
\begin{center}
\includegraphics[width=6cm, angle=-90]{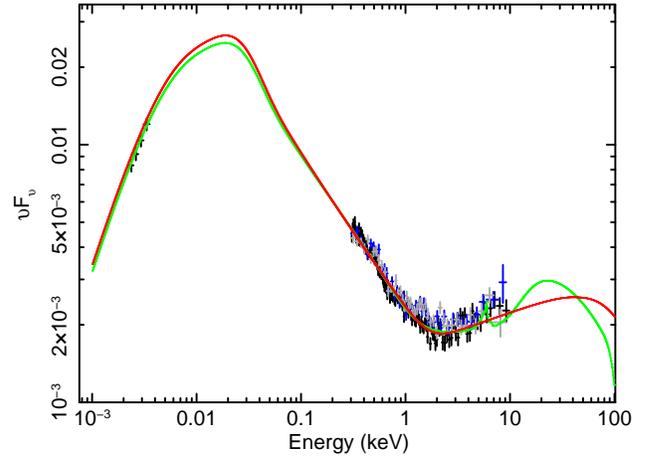}
\caption{Unabsorbed SED at 2001 October (black points optical+pn, blue points MOS1, grey points MOS2), with the colour temperature corrected accretion disc spectrum with Comptonisation, {\it optxagnf} (red curve), and including reflection, {\it optxagnf+(rdblur*pexmon)} (green curve). The y-axis is in units keV$^2$ photons cm$^{-2}$ s$^{-1}$ keV$^{-1}$ and energies are in the observer frame.}
\label{Modelsand2001data}
\end{center}
\end{figure}

\subsection{Epoch 2001}
We began by modelling the SED taken in 2001, when the source flux was at its highest. The data at this epoch comprise {\it XMM-Newton} EPIC pn and MOS, and optical spectroscopy in the blue grating of the 1.9m Radcliffe telescope.

We adopted the physical model {\it optxagnf} of D12. This comprises a colour temperature corrected accretion disc spectrum with Comptonisation of both low temperature optically thick disc material and high temperature optically thin material in a corona. Black hole mass was fixed at 10$^{8.1}$ M$_{\odot}$ from the H$\beta$ broad-line measurement (Section \ref{sec:bhmass}), and distance fixed at $z=0.1015$, again as measured from the optical spectra. 
We also fixed the black hole spin to zero and outer disc radius to 10$^5$\,R$_{\rm g}$. 
In addition to the fixed Galactic column, we allowed for an intrinsic absorber ({\it tbabs}, \citealt{Wilms}) and correlated reddening $E(B-V)_{\rm int} = 1.7 \times N_{\rm H,int}$/10$^{22}$ \citep{Bessell}.

Fitting this model to the data using $\chi^2$ minimisation we found the best-fitting power law slope to be $\Gamma=1.87$. The resulting black hole accretion rate was $L/L_{\rm Edd} = 0.27$ (resulting in $\chi^2$/$\nu$ = 1172/1058). The parameters resulting from this fit are very similar to those of one of the mean AGN spectra presented in J12a, Jin, Ward \& Done 2012b and D12 in which $\langle M/M_{\odot} \rangle~\sim 10^8$ and $\langle L/L_{\rm Edd} \rangle~\sim 0.2$. Our measured $L/L_{\rm Edd}$ is consistent with that measured by \cite{Gru10} from {\it Swift} data alone and using different models. Remarkably, no intrinsic absorption was required at all.

At energies above 2 keV, where our data are less constraining, this model assumes a pure power law. Many AGN are, however, well fitted with reflection off the disc above 2 keV, so we went on to include a reflection component with the {\it pexmon} Xspec routine \citep{pexmon}. We assumed fixed solid angle $\Omega/2\pi = 1$, inclination angle 30$^{\circ}$ and $\Gamma$ tied to the photon index of the disc+Comptonisation model, using {\it rdblur} (based on \citealt{Fabian89}) with $R_{\rm in} = 15$\,R$_{\rm g}$ to smear the Fe~K$\alpha$ line since no narrow emission lines are apparent at this epoch. The total disc+Comptonisation+reflection model is a good representation of the broadband data ($\chi^2$/$\nu$ = 1156/1058), again with no measurable intrinsic neutral absorption. The power law photon index steepens a little from $\Gamma = 1.87$ to 2.07 when reflection is included, while the soft X-ray components and inferred accretion rate remain approximately the same ($L/L_{\rm Edd} = 0.25$). The Comptonisation component responsible for the soft X-ray excess here is optically thick. Both models and their contributing components are shown in Figure \ref{Modelcomponents}, and the data are shown with these models in Figure \ref{Modelsand2001data}. 
The best fits for both models are listed in Table \ref{tab:sedparams}. These can be compared with Table 3 of D12, and appear to be similar to their mean AGN model M2. 

We note that, while we have no reason for allowing an offset between the optical and X-ray, if we do that no dramatic changes in the remaining free parameters are required and we obtain a comparable fit statistic. The optical constant factor goes to 0.8.
\begin{table*} 
\begin{center}
\caption{Overview of the best-fitting {\it optxagnf[+(rdblur*pexmon)]} SED model parameters for epoch 2001. Not listed are the fixed (frozen) parameters $z = 0.1015$, $M = 10^{8.1}$\,M$_{\odot}$, $d_l = 462.1$\,Mpc, $N_{\rm H,Gal} = 9.43 \times 10^{19}$\,cm$^{-2}$, $E(B-V)_{\rm Gal} = 0.016 = 1.7 \times N_{\rm H,int}$/10$^{22}$ magnitudes. $^*$ the two photon indices were tied between {\it optxagnf} and {\it pexmon}. The large number of free parameters relative to datapoints here makes obtaining explicit error bars on each parameter a difficulty, therefore we do not present errors here but the uncertainties can be seen, in the broad sense, by comparison of the two models we have fitted.}
\begin{tabular}{lllll} 
Model & parameter & unit                      & {\it optxagnf}  & {\it optxagnf+(rdblur*pexmon)} \\ \hline \hline
  $zTBabs$     &$N_{\rm H}$ &10$^{20}$ cm$^{-2}$   & 0 & 0\\
  $zredden$    &$E(B-V)$     & =1.7x$\frac{N_{\rm H}}{10^{22}}$ mag & 0 &0\\
  $rdblur$     &Index      &        & -    & 3 fixed\\
  $rdblur$     &$R_{\rm in}$     &   R$_{\rm g}$&     -& 15 fixed\\
  $rdblur$     &$R_{\rm out}$    &  R$_{\rm g}$     &-  &400 fixed   \\
  $rdblur$     &Incl       &deg    &-  &30 fixed     \\
  $pexmon$     &$\Gamma^*$   &      &-   &2.07$^*$     \\
  $pexmon$     &$E_{\rm fold}$      &keV   &-  & 1000 fixed    \\
  $pexmon$     &rel\_refl   &      &-  & -1 fixed ($|$R$|$=1)  \\
  $pexmon$     &abund     &     &-     &1 fixed  \\
  $pexmon$     &Fe\_abund  &      &-   & 1 fixed   \\
  $pexmon$     &Incl     & deg    &-  &30 fixed   \\
  $pexmon$     &norm     &       &-    &0.002 fixed\\
  $optxagnf$   &log($L/L_{\rm Edd}$) &    & -0.57 & -0.61  \\
  $optxagnf$   &astar     &         & 0 fixed       &0 fixed \\
  $optxagnf$   &$R_{\rm cor}$       &  R$_{\rm g}$     & 34     & 33 \\
  $optxagnf$   &log(R$_{\rm out}$)    & R$_{\rm g}$ & 5.0 fixed &5.0 fixed \\
  $optxagnf$   &$kT_e$       &keV     & 0.28   & 0.21 \\
  $optxagnf$   &$\tau$       &       &   13   &  15 \\
  $optxagnf$   &$\Gamma$     &       & 1.87&  2.07$^*$  \\
  $optxagnf$   &$f_{\rm pl}$       &        &  0.30    & 0.36 \\
  $optxagnf$   &norm       &       &  1.0 fixed     &1.0 fixed \\\hline
$\chi^2$/$\nu$ &        &        &1172/1058& 1156/1058 \\
\label{tab:sedparams}
\end{tabular}
\end{center}
\end{table*}

The distinction between these models lies primarily above a few tens of keV. The 15--50\,keV {\it Swift} BAT data do not show any detection, even in the averaged flux considered for the BAT Survey \citep{Baum}. The upper limit on its 70-month averaged flux is approximately 1\,mCrab, which does not allow discrimination between these models.

\begin{figure}
\begin{center}
\includegraphics[width=6cm, angle=-90]{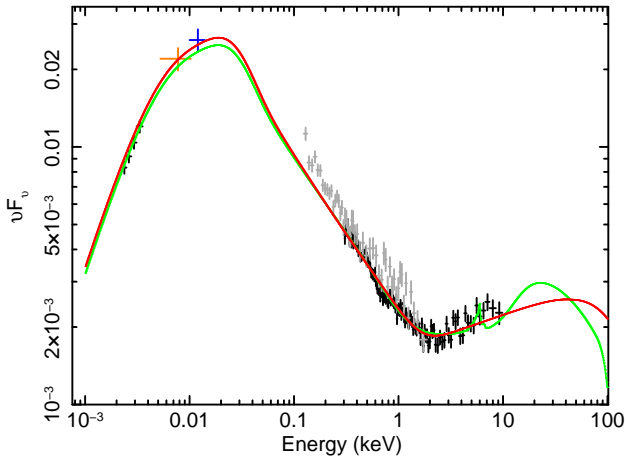}
\caption{Unabsorbed SED at 2001 October (black points, showing optical and EPIC pn only for clarity), with the two possible models. We overplot the archival unabsorbed spectrum from {\it ROSAT} PSPC (grey) and the approximate continuum fluxes observed with IUE (orange) and FUSE (blue) assuming 10 per cent errors on each. The y-axis is in units keV$^2$ photons cm$^{-2}$ s$^{-1}$ keV$^{-1}$.}
\label{archivaldata}
\end{center}
\end{figure}
To extend the energy coverage towards the peak of the $\nu F_{\nu}$
SED we overplotted the archival {\it ROSAT} PSPC spectrum
\citep{aad,emp95}, and de-absorbed this according to the 2001 spectrum
with extrapolation using our {\it optxagnf} model (Figure
\ref{archivaldata}). The soft X-ray spectrum in the PSPC data shows
that the source had a strong soft excess in 1993 which is consistent
with the 2001 data (it appears a little softer, which has been noted
previously for PSPC spectra - see Section 2.3.1 of \citealt{aad} for
further details). We also estimated the approximate UV continuum flux
in the {\it IUE} SWP \citep[taken in 1992,][]{DunnIUE} and combined
{\it FUSE} \citep[taken 2002--2004,][]{DunnFUSE10} observations using
published spectra, and de-absorbed these using the LAB Survey Galactic
value. Both are consistent with a UV spectrum rising to shorter
wavelengths (within the 10 per cent errors we have assumed in Figure
\ref{archivaldata}), but we caution that source flux estimation in the
UV is heavily absorption-model dependent.

We note that \cite{emp95} performed SED fitting of the 1993 PSPC
spectrum with the {\it ROSAT} WFC point and an optical spectrum taken
2 years previously with the SAAO 1.9m Radcliffe telescope, and also
found an optical continuum rising to the blue and forming a big blue
bump with the soft X-ray data.

Our model gives a monochromatic, unabsorbed luminosity at 200\,eV of
$\log L(200)=45.06$, somewhat lower than the WFC luminosity of $\log
L(200)=45.61$ \citep{Edelson}. Some of this discrepancy is due to
the larger Galactic absorption column used \citep[$1.4\times 10^{20}$ compared to our $0.93\times 10^{20}$~cm$^{-2}$,][]{Edelson}. However, this
only increases our $\log L(200)$ to $45.22$, so there may be some
intrinsic variability such that the source was brighter during the WFC
survey than in our dataset.

\begin{figure}
\begin{center}
\includegraphics[width=6cm, angle=-90]{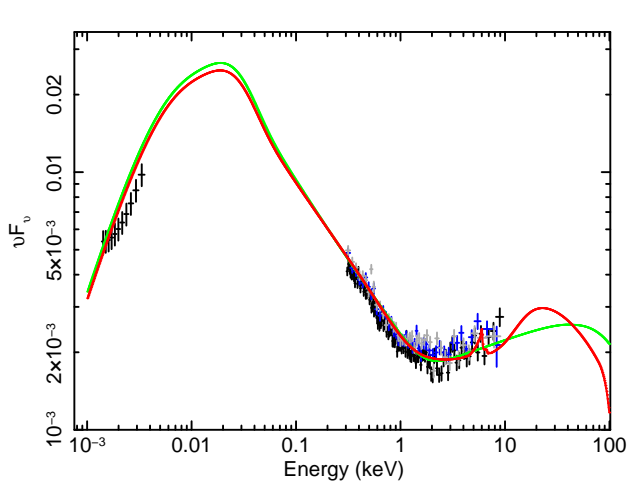} 
\caption{SED of October 2000 (black, blue and grey points for pn+optical, MOS1 and MOS2) with the epoch 2001 models overlaid (solid red and green lines as before). The y-axis is in units keV$^2$ photons cm$^{-2}$ s$^{-1}$ keV$^{-1}$.}
\label{2000data2001model}
\end{center}
\end{figure}
\subsection{Epoch 2000}
\begin{figure}
\begin{center}
\includegraphics[width=6cm, angle=-90]{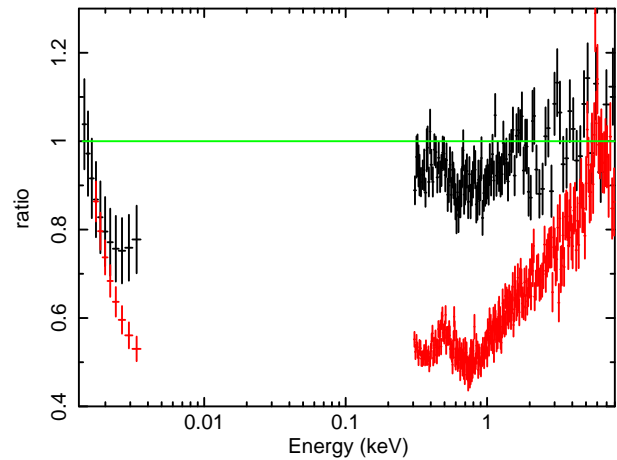} 
\caption{Ratio of the SEDs in October 2000 (black) and May 2007 (red) to the 2001 Model 1. For clarity we plot only the rebinned EPIC pn + optical data.}
\label{ratio}
\end{center}
\end{figure}

The model parameters which describe the epoch 2001 SED are also a reasonable representation of the epoch 2000 SED (comprising EPIC pn, MOS1+2 and optical spectra), as shown in Figure \ref{2000data2001model}. We do, however, have a small flux deficit in optical and soft X-rays (Figure \ref{ratio}), and this can be accounted for most simply by additional intrinsic X-ray absorption (using {\it tbabs}) of $N_{\rm H} = 1 \times 10^{20}$ cm$^{-2}$, and additional intrinsic optical reddening of $E(B-V) = 0.07$ mag. The resulting fit statistic for the {\it optxagnf} model is $\chi^2$/$\nu$ = 1307/1071, and we note that to achieve this fit we untied the optical extinction and it converged at a value much greater than $1.7 \times N_{\rm H}$. The key point here is that it is possible to keep the same continuum as seen in 2001, and change only the absorption to describe the 2000 SED. We also tried fixing all parameters but the power law photon index and we obtain a good fit to the X-ray data with a flatter $\Gamma = 1.81$, while the optical data are still overpredicted by the model and additional reddening must be included.

\begin{figure}
\begin{center} 
\includegraphics[width=6cm, angle=-90]{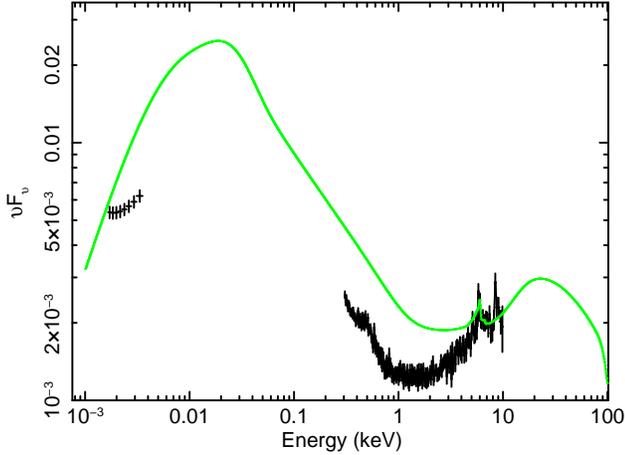}
\caption{SED of May 2007 (black points) with the epoch 2001 {\it optxagnf+(rdblur*pexmon)} model overlaid (solid green line). For clarity we plot only the rebinned EPIC pn + optical data and y-axis is in units keV$^2$ photons cm$^{-2}$ s$^{-1}$ keV$^{-1}$.}
\label{epoch2007with2001model}
\end{center}
\end{figure}
\begin{figure*}
\begin{center}
\includegraphics[width=6cm, angle=-90]{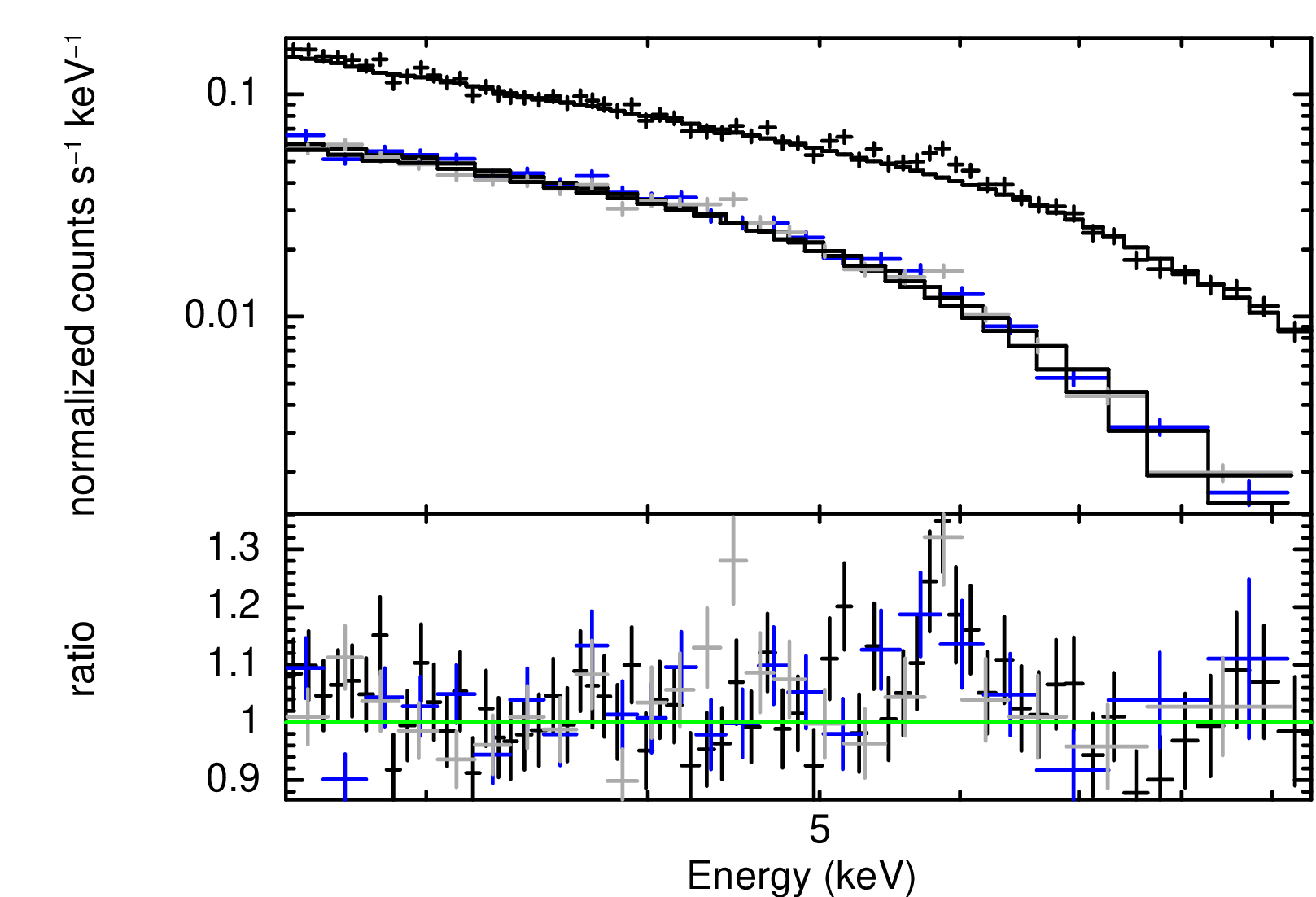}
\includegraphics[width=6cm, angle=-90]{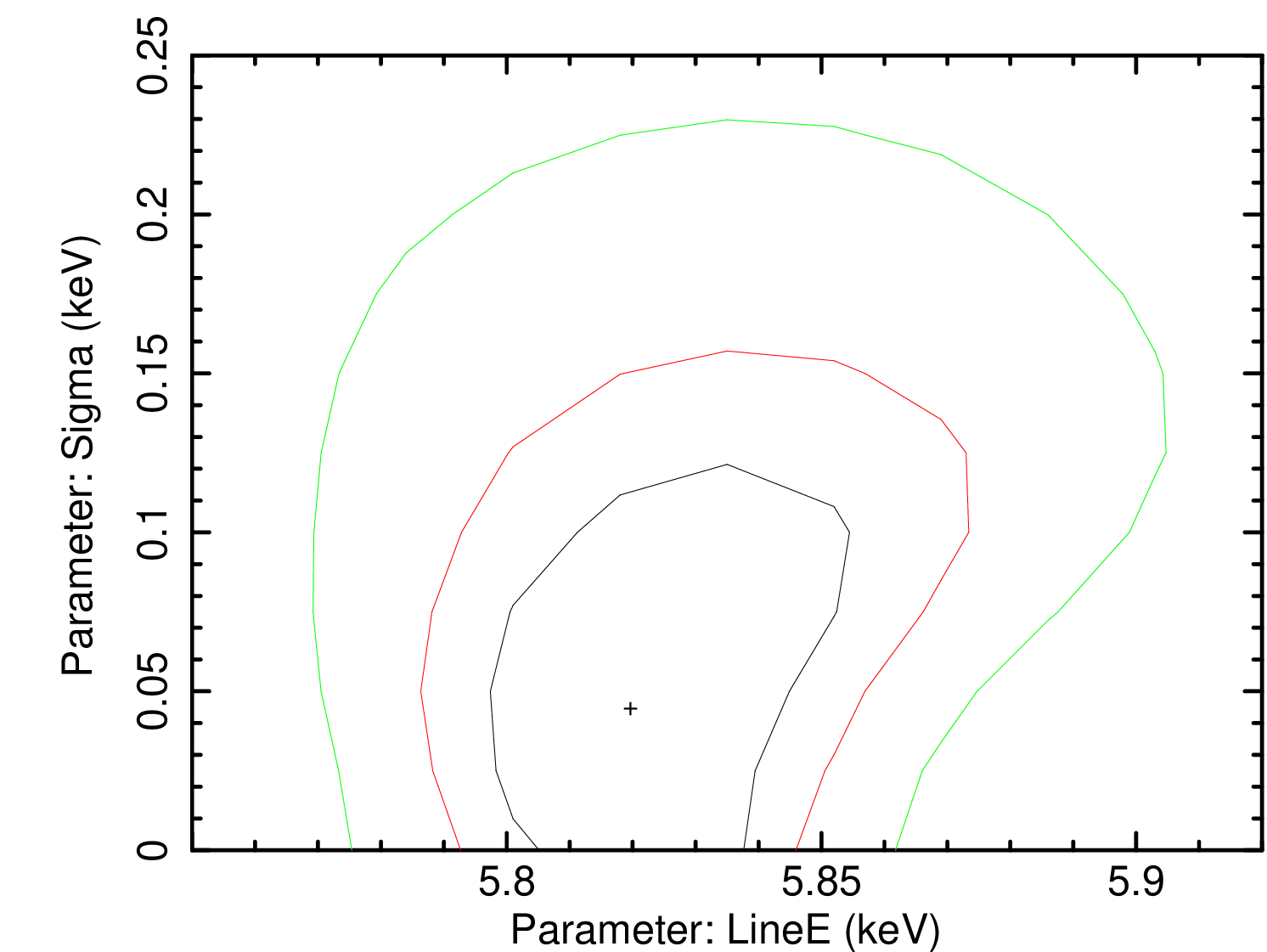}
\caption{Left: Data (upper panel) and data to model ratio (lower panel) for the 2.5--9.5\,keV XMM data from 2007. A power law model and the pn data are shown in black, MOS1 in blue and MOS2 in grey. An excess at the expected energy of an Fe line is apparent. Right: Contour plot showing the 68, 90 and 99 per cent confidence contours on the observed Gaussian line energy and width sigma.}
\label{Feline}
\end{center}
\end{figure*}

\subsection{Epoch 2007}
A simple overlay of the models adopted for the 2001 epoch is a poor description of the SED in 2007 (comprising EPIC pn, MOS1+2, Swift XRT and optical data), as can be seen in Figure \ref{epoch2007with2001model}. An event must have occurred that suppressed both the bluemost optical and soft X-ray flux, yet left the hard ($>$\,5\,keV) X-ray and redmost optical flux unaltered. The sharp features seen in residuals are strongly suggestive of absorption (Figures \ref{ratio} and \ref{epoch2007with2001model}).

The 2007 {\it XMM-Newton} observation was the longest X-ray
observation of RE\,J2248-511 with exposure of almost 60\,ks. In these
data an Fe line is detected, and fitting a Gaussian we measure an observed line centre of
5.82$\pm$0.04\,keV (i.e. rest frame 6.4~keV) and line width $\sigma \le 0.15$\,keV as shown in Figure \ref{Feline}. This leads us to prefer a model for the continuum which includes reflection. 

Maintaining the same underlying disc+Comptonisation+reflection continuum, and fitting for additional intrinsic neutral absorption ({\it tbabs}) and reddening ({\it E(B-V)}) did not result in an acceptable fit and the SED shape was not well fitted ($\chi^2$/$\nu$ = 147538/2049). We then added a partial covering neutral absorber, {\it pcfabs}, and left {\it E(B-V)} and {\it tbabs} $N_{\rm H}$ at the source free to vary independently, and obtained a much improved fit. The resulting partial covering column density required was $N_{\rm H} = 5.8 \times 10^{22}$ cm$^{-2}$ with covering fraction, $f_{\rm cov}$, 0.4, and the neutral, fully covering absorbing column remained small (see Table \ref{tab:abs}). This was the simplest absorption model that could explain the large drop in optical to soft X-ray flux between 2001 and 2007, although it is likely that any absorption is far more complex in nature. Replacing the fully covering neutral absorber with an ionised absorber we found that the depth of the Fe Unresolved Transition Array \citep[e.g.][]{Sako} was greatly overpredicted. 

We examined data from the high resolution {\it XMM-Newton} Reflection Grating Spectrometer to shed further light on the details of the X-ray absorber. Unfortunately, there was insufficient signal to measure any spectral features.

\begin{table} 
\begin{center}
\caption{Intrinsic absorption required at epoch 2007, in addition to the {\it optxagnf+(rdblur*pexmon)} model parameters best-fitting at epoch 2001 (listed in Table \ref{tab:sedparams}).}
\begin{tabular}{lllll} 
Model & parameter & unit & value \\ \hline \hline
  $zTBabs$     &$N_{\rm H}$ &10$^{20}$ cm$^{-2}$   & 1.0\\
  $zredden$    &$E(B-V)$     & =1.7x$\frac{N_{\rm H}}{10^{22}}$ mag & 0.11\\
  $pcfabs$     &$N_{\rm H}$ &10$^{22}$ cm$^{-2}$   & 5.8\\
  $pcfabs$     &$f_{\rm cov}$ & &  0.4\\ \hline
$\chi^2$/$\nu$ &        &        & 3038/2049\\
\label{tab:abs}
\end{tabular}
\end{center}
\end{table}

\subsection{Epoch 2006}
\begin{figure}
\begin{center} 
\includegraphics[width=6cm, angle=-90]{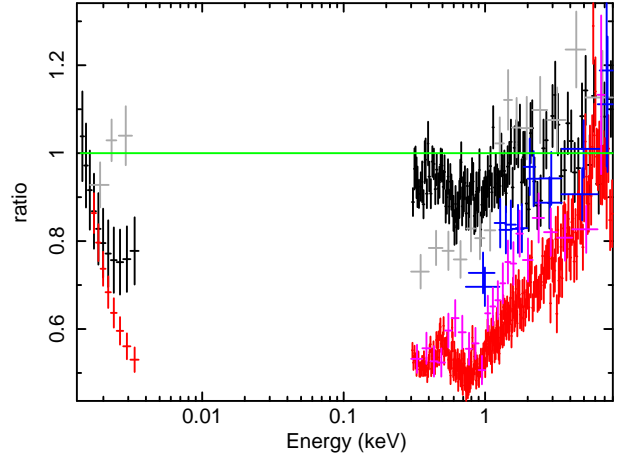}
\caption{Ratio of the SEDs in October 2000 (black, pn+optical), May 2007 (red, pn+optical), September 2007 (pink, {\it Swift} XRT), September-November 2006 (grey, XRT+optical) and 1997 (blue, {\it ASCA} GIS) to the best-fitting 2001 {\it optxagnf} model.}
\label{ratioall}
\end{center}
\end{figure}

We fitted the September-November 2006 SED last because neither {\it XMM-Newton} data nor optical spectroscopy were available. The {\it Swift} XRT X-ray spectrum is similar in shape to that of 2007 with some small increase in flux (Figure \ref{ratioall}). We can therefore confirm that the strong X-ray absorption seen in 2007 was also present in 2006, to a lesser extent. The X-ray data are not of sufficient signal-to-noise to discriminate between differing models, however, nor to extract details of either the continuum or the absorption model. 

The optical photometry, reported in Section \ref{sec:obs}, includes a contribution from the emission lines. We estimated and removed the line contributions using previous optical spectra. The resulting $B$ and $V$ band fluxes are higher than measured in 2007 via spectroscopy, while the $R$ band flux is approximately the same. When plotted with the X-ray data in $\nu F_{\nu}$ space the optical continuum shows little deviation from the 2001 model in contrast to the X-ray spectrum at this epoch (Figure \ref{ratioall}), but we caution that the errors may be underestimated if there are significant systematic uncertainties.

Finally, we overplotted the archival {\it ASCA} GIS spectra \citep{aad} taken in 1997, and find that these lie between the 2001 and 2007 spectra examined here (Figure \ref{ratioall}).

\subsection{Long term variability: summary of all epochs}
These data allow the long-term spectral shape to be assessed at 6
epochs between 1992 and 2007. In 1993, the {\it ROSAT} spectrum
agrees well with our model for the 2000 and 2001 data, as do the 1992
{\it IUE} and 2002--2004 average {\it FUSE} continua (with the caveat
of strong absorption model dependency), suggesting the source was not
significantly absorbed at these epochs. The {\it ASCA} spectra, taken
in 1997, cover the portion of the spectrum dominated by Compton
upscattering and reflection in our model. The fit residuals show a
decline towards softer energies when compared with the 2001 model. In
particular the lowest GIS energy bins covering 0.7--1.0 keV show
0.65--0.8 times the flux seen in 2001, but this is not conclusive and
without coverage of soft X-ray or optical/UV regimes we cannot infer
the presence or absence of absorption. In 2006 the XRT spectrum lies
below the 2001 data, and we can infer that absorption is present when
assuming the 2001 best-fitting model. In 2007 the XRT, EPIC and
optical spectra show a large deficit in flux below 5\,keV which we
interpret as absorption.

\section{Discussion} \label{sec:discussion}
The optical to X-ray SED of RE\,J2248-511 can be well represented by a
colour temperature corrected accretion disc spectrum with
Comptonisation of both low temperature optically thick disc material
and high temperature optically thin material in a corona, plus a
reflection component which is evident above 5\,keV. In 2000--2001
little or no intrinsic absorption is required. The detection of this
object in the FUV by the {\it ROSAT} WFC similarly requires that there was
little or no intrinsic absorption at this epoch (1990) also. 

However, the spectrum dramatically altered over the space of 5 years
in the soft X-ray and optical/UV, displaying a reduction in flux in
both optical and X-ray, strongly suggestive of absorption. Indeed,
additional fully and partially covering neutral absorbers plus
additional reddening are able to explain the difference in SED shape
seen in 2006--2007 whilst preserving the underlying continuum as seen
in 2000--2001.

We know from detailed spectroscopic UV observations with {\it IUE}, {\it FUSE} and {\it HST} COS that complex absorbers exist in this system \citep{DunnIUE,DunnFUSE08,DunnFUSE10,BorguetHST}. These appear to lie at extremely large distances, (9-15\,kpc from the AGN) i.e. out into the halo of the host galaxy, so they possibly represent a galactic wind. Thus their density is extremely low, so they cannot vary in response to the changing illumnation from the AGN, and indeed these are observed to remain constant.

The X-ray absorption we find here went unnoticed in the aforementioned
work by \cite{DunnFUSE10}. These authors looked only at the 2007 {\it
XMM-Newton} spectrum, and were therefore unable to detect variability
in the spectral shape and model any constant, underlying
continuum. They concluded the X-ray spectrum was not absorbed, and
compared this with the {\it FUSE} UV spectra taken in 2002 and 2004
which show evidence for significant absorption. From this they
suggested that the source does not follow the 1:1 correspondence
between UV and X-ray absorption claimed for low redshift AGN
\citep{Crenshaw,Kriss}. Our findings show, however, that
RE\,J2248-511 is strongly absorbed in the optical/UV to X-rays at some
epochs and indeed these appear to be partially correlated.

A plausible scenario is that an absorbing cloud (system) crossed into
our line-of-sight to the central AGN between 2005 and 2007,
dramatically altering the optical to X-ray SED. The neutral X-ray
absorber required an increase in column density of $\sim$10$^{20}$
cm$^{-2}$, plus $\sim$0.3 magnitudes of optical extinction ($A_V$),
while an additional 40\% partially covering X-ray absorber was
invoked, with a large column of $\sim$5$\times$10$^{22}$ cm$^{-2}$ if
neutral. This could be the onset of a new outflow, or a discrete
episode of mass ejection. It is possible that a similar event occurred
around 1997 as viewed by {\it ASCA} \citep{aad}, or at least occurred
some time in between the epochs of 1993 and 2000 when strong soft
excesses are clearly present; this is possibly also the case in 1992
when a flat optical continuum was noted \citep{Ma} in contrast both to
the bluer spectrum observed in 1991 \citep{Gru98,Gru} and large soft
X-ray excess seen with {\it ROSAT} in 1993 \citep[][see also Breeveld
et al. 2001]{emp95}.

To put RE\,J2248-511 in the wider context, we compare it to sources among the similarly analysed sample presented in J12a,b,c. J12a took an X-ray/optically selected sample of
unobscured {\it XMM-SDSS} Type I AGN, and modelled the SEDs with {\it
optxagnf} as we have done here. Their sample contained 51 AGN, among
which $\sim$20\% were NLS1. They concluded that these AGN could be
carved up into 3 SED types and they reported the main parameters for
each. The 2001 SED for RE\,J2248-511 is entirely consistent with their mean SED
which has $\langle M/M_{\odot} \rangle~\sim 10^8$, $\langle L/L_{\rm
Edd} \rangle~\sim 0.2$, $\langle R_{\rm cor} \rangle~\sim 40$,
$\langle kT_{\rm e} \rangle~\sim 0.3$, $\langle \tau \rangle~\sim 13$,
$\langle \Gamma \rangle~\sim 1.87$ and $\langle f_{\rm pl}
\rangle~\sim 0.3$. The Eddington fraction is perhaps the most likely
parameter to
be driving the spectral shape, and while NLS1s are around Eddington
in most samples, we find a value of $L/L_{\rm Edd} \sim
0.25-0.27$, confirming RE\,J2248-511 as a typical BLS1s
\citep[e.g.][J12a]{VandF09,Gru10}.

RE\,J2248-511 lies in a direction of low Galactic absorption. Combined
with the total lack of measurable intrinsic absorption and reddening,
this means we are seeing the AGN through a `hole' in the Galactic and
host galaxy column density. The continuum flux, which peaks at EUV
energies, arrives uninhibited and as a consequence RE\,J2248-511
appears in EUV-selected samples. Given the very typical BLS1 continuum
model parameters we obtain when fitting the non-absorbed epoch SEDs,
this explanation is very attractive. This suggestion was originally
made by \cite{emp95}, but with the potential for such dramatic
changes caused by absorption alone, it is only with multi-epoch,
multi-wavelength data that we can now confirm this. The {\it ROSAT}
WFC Extragalactic Survey AGN sample \citep{Edelson} in which
RE\,J2248-511 was catalogued, contains 4--5 AGN among the sample of 19
(including BL Lacs) with Galactic columns $<10^{20}$ cm$^{-2}$.
The low intrinsic column and {\it ROSAT} WFC, {\it IUE} and {\it FUSE} detections show that there can be a high escape fraction for EUV and FUV photons from such AGN, which can be important for the AGN contribution to re-ionisation of the Universe, both in terms of hydrogen and helium.

In summary, RE\,J2248-511 is a high black hole mass, broad-lined AGN
in the local Universe, with an Eddington ratio similar to broad-lined
quasars. Our modelling demonstrates that this spectral shape can be
accommodated with a colour temperature corrected accretion disc
spectrum with Comptonisation of both low temperature optically thick
disc material and high temperature optically thin material in a
corona, plus a reflection component. On timescales of a few years the
soft excess shows dramatic variability, which can be readily explained
by the onset of absorption from both fully and partially covering
material. We conclude that rather than an unusual, ultrasoft AGN which
defies classification, RE\,J2248-511 is in fact a typical broad-lined
Seyfert 1 that is fortuitously viewed through a `hole' in the line-of-sight column
density.

\section{Acknowledgments}
RLCS acknowledges financial support from a Royal Society Dorothy Hodgkin Fellowship. We thank Jens Hjorth and Johan Fynbo for generously giving us observing time at the Danish Telescope, and Francois van Wyk for assistance during our first observing run at SAAO. We thank Hans Krimm and Wayne Baumgartner for assistance with the {\it Swift} BAT data, and Steve Sembay and Roberto Soria for useful discussions. KW acknowledges support from STFC. KLP acknowledges support from UKSA.
This work is based on observations obtained with \xmmn, an ESA science mission with instruments and contributions directly funded by ESA Member States and the USA (NASA).
This work made use of data supplied by the UK {\it Swift} Science Data Centre at the University of Leicester, and data provided by the High Energy Astrophysics Science Archive Research Center (HEASARC), which is a service of the Astrophysics Science Division at NASA/GSFC and the High Energy Astrophysics Division of the Smithsonian Astrophysical Observatory. This paper uses observations made at the South African Astronomical Observatory (SAAO). We acknowledge use of the AAVSO Photometric All-Sky Survey (APASS), funded by the Robert Martin Ayers Sciences Fund. IRAF (Image Reduction and Analysis Facility) is distributed by the National Optical Astronomy Observatories, which are operated by AURA, Inc., under cooperative agreement with the National Science Foundation. 

We dedicate this work to the memory of Liz Puchnarewicz, who brought this source to the fore and inspired its further study.

\bsp

\label{lastpage}

\end{document}